# A Governance Methodology Layer for AI-Assisted Software Development: Defect Taxonomy, Controlled Ablation, and a Test of Process-Over-Capability

권성진 (Kwon Sungjin)

*Independent Researcher · akaa1941@gmail.com*

## Abstract

Autonomous coding agents produce output that passes syntactic checks — compilation, type safety, CI — at high velocity. Yet syntactic correctness does not imply semantic correctness: design boundaries, security invariants, and maintainability contracts remain structurally invisible to automated pipelines. This paper makes four contributions. First, we present a **defect-class taxonomy** grounded in five AI agent permission and governance modules, separating defects structurally detectable by static analysis from those requiring semantic review. Second, we describe a **runtime-decoupled governance gate** — a file-based protocol that reads generator output and emits a structured verdict without API coupling, hence portable across generators. Third, we formalize **methodology-as-code**: expressing a verification protocol as a version-controlled, executable artifact whose two layers separate portable methodology from host automation. Fourth, we report a **controlled ablation experiment** (E-ablation, N=5 artifacts, 8-item independent ground truth) comparing harness-structured review against a token-matched unstructured prompt. The structured condition records 62% lenient recall against 50%, with 25% strict against 0%. A severity-grade differential reported earlier does not survive blind re-grading and is withdrawn (§6.6). An independent-session re-test with blind scoring does not replicate that contrast: the conditions differ by one strict hit in 24 (6/24 against 5/24), the structured aggregate again 25% and the unstructured 0% not recurring (§6.7). Both conditions miss document-quality defects identified by a human QA reviewer, indicating complementarity between structured AI review and human process inspection. The re-test does not distinguish structured review from a detailed unstructured prompt here, and a later pre-registered test that also varied model tier (184 runs, §6.8) did not support **process design as the dominant factor**: under strict scoring the tier effect was the larger one, and the structured prompt did not outperform a one-sentence prompt.

## 1. Introduction

### 1.1 The Semantic Gap in AI-Assisted Development

Modern coding agents — OpenCode, GitHub Copilot, Devin, Claude Code — substantially reduce implementation time. The automation pipeline (specification → generation → CI → merge) appears to close the development loop. But it closes only the *syntactic* loop. The semantic loop — design coher-ence, security boundary enforcement, permission correctness, maintainability contract — remains structurally open for three reasons:

1. **CI is syntactic, not semantic.** Compilation, type checking, and unit tests verify that code is well-formed and that specific input-output pairs hold. They do not verify that a permission allowlist covers all execution paths, that a session-scoped grant is not silently persisted as permanent, or that an LLM trusted as the sole security gate has no deterministic backstop.
2. **Same-model self-review is structurally biased.** The model that generates code shares the same training distribution as the model that self-reviews it. Blind spots are symmetric: the same reasoning failure that produced a defect is likely to miss it during review.
3. **Velocity incentivizes shallow gates.** At agent velocity, a human reviewer faces dozens of diffs per day. The marginal value of deep review declines as volume rises. A governance layer must provide depth without proportional human time.

This paper investigates two questions. First, whether a model-agnostic, version-controlled governance methodology narrows the semantic gap described above. Second, whether the methodology structure is itself the operative factor, or whether any sufficiently detailed review prompt yields equivalent coverage. The second question is treated as the primary rival explanation and is tested directly in the controlled ablation of §6.

### 1.2 This Paper's Scope

This paper is deliberately distinct from our companion work on governance economics (v2 paper). That work quantifies the *governance dividend* — coverage gain per token, multi-model dynamics, and a model-tier comparison — and is framed as an AI systems paper. This paper is an SE paper: the subject is the **methodology as a software artifact**, and the central claim is about **process design**, alongside model capability.

Specifically, we contribute:

- **C1** (§3): A defect-class taxonomy grounded in five real AI agent modules
- **C2** (§4): A runtime-decoupled governance gate architecture
- **C3** (§5): Methodology-as-code and the 2-layer portability pattern
- **C4** (§6): A controlled test of whether process design, rather than model capability, drives coverage. The contrast this test reports does not survive an independent re-test under condition-blind scoring (§6.7), and a pre-registered test that also varied model tier did not support it (§6.8); this paper does not show that process design outweighs model capability.

### 1.3 Relation to Prior Work

Our v1.0 paper argued structurally that a harness is worth building — a durable configuration layer atop LLM-based tools. Our v2 paper measured the governance dividend empirically across five FH-in-ternal artifacts. This paper introduces an **external artifact set**, an **independent ground truth**, and the **controlled ablation** that prior work flagged as the highest-priority future experiment.

---

## 2. Background

### 2.1 The Syntactic-Semantic Correctness Gap

Prior SE literature (Fagan 1976; Myers 1978; Wiegers 2002) establishes that structured human inspection catches defect classes that automated testing misses. The defect classes differ: automated testing finds *behavioral* defects (wrong output for a given input), while inspection finds *design* defects (wrong contract, wrong boundary, wrong invariant). This distinction predates LLMs.

What changes with LLMs is velocity and volume. When a human writes 200 lines/day, a team can afford deep review. When an agent writes 2,000 lines/day, the same review process breaks. The SE challenge is not to replace testing — testing still catches behavioral bugs — but to make semantic review scalable.

### 2.2 AI Agent Permission and Governance Code

AI coding agents operate in environments with rich tool access: file system, shell, network, databases. The boundary between "what the agent is allowed to do" and "what it does" is enforced by permission and governance modules: arity tables, permission judges, approval callbacks, memory freshness validators. These modules are themselves AI-adjacent — some are AI-generated, all are AI-facing — and their semantic correctness is critical to the security posture of the entire system.

Our artifact set is drawn from this domain: permission enforcement code in active AI agent frameworks (OpenCode, Hermes, Goose, OpenHuman). This is not incidental. If governance methodology catches defects in governance code itself, it is self-validating in the most demanding domain.

### 2.3 Structured vs. Unstructured Review

The closest prior work on structured vs. unstructured AI review is in the prompt engineering literature (Wei et al. 2022; Yao et al. 2023), which shows that structured prompting (chain-of-thought, role assignment, step decomposition) improves performance on reasoning tasks. However, this literature evaluates output *quality* on tasks with ground truth, not *defect recall* against an independently established checklist. The E-ablation experiment of §6 addresses this gap for the software review domain.

### 2.4 The Natural-Language Harness as a Research Object

The methodology described in this paper sits at the intersection of two communities. To the AI systems community (cs.AI), the central artifact is a natural-language object interpreted by an LLM runtime; to the software engineering community (cs.SE), it is a version-controlled process specification subject to the software development lifecycle. This paper takes the SE view, but the underlying object is shared, and recent AI-systems work has begun to study it directly.

Pan et al. (2026) introduce *natural-language agent harnesses* (NLAH), in which the control logic of an agent harness — normally embedded in code — is externalized as an executable natural-language specification interpreted at runtime, and then ablated to measure which components affect task resolution. Their work provides external empirical support for the premise underlying the present paper: that the design-pattern layer of a harness can be expressed as a natural-language artifact rather than as code. The convergence is on the artifact form; the contributions differ in objective. NLAH measures the form — whether the natural-language harness migrates across tasks and which modules move task-resolution rates. The present paper governs and compounds the form — it treats the natural-language specification as a software artifact subject to adversarial verification, drift control, and cross-session accumulation, and measures defect-coverage rather than task-resolution. The two lines are therefore complementary: one establishes that the form is measurable and effective, the other addresses how such a specification is verified and maintained as a quality layer over time.

---

## 3. Defect-Class Taxonomy (C1)

### 3.1 The Structural Argument

CI and static analysis tools operate on the syntactic structure of code. A linter checks style and obvious bugs. A type checker verifies type consistency. A SAST tool scans for known vulnerability patterns (SQL injection, XSS, hardcoded credentials). None of these tools ask: *does the permission boundary correctly reflect the intended security model?* or *does this function's error contract match how callers use it?*

We identify four defect classes that are structurally beyond CI/SAST reach:

| Class | Definition | Example |
|---|---|---|
| **D1 — Semantic boundary** | A logical invariant that is correct syntactically but violates intended semantics | `allow_session` granted via `kind="allow_always"` — type-correct, semantically wrong |
| **D2 — Silent failure mode** | An error path that produces a valid-typed output masking an exceptional condition | `except Exception: return "deny"` — caller cannot distinguish error from legitimate denial |
| **D3 — Missing contract guard** | A precondition documented in comments/types but not enforced | `ageDays >= 0` as a type comment, clamped silently at three independent sites |
| **D4 — Dead correctness code** | Guards that are mathematically unreachable given input constraints | `recall < 0` after `2^(-age/halfLife)` where `age >= 0`, `halfLife > 0` |

### 3.2 Grounding in Artifact Evidence

We grounded this taxonomy in five AI agent permission and governance modules reviewed in §6 (E-ablation). Across the five artifacts, the structured review condition found 49 defects, of which:

- **D1 (Semantic boundary)**: 11 findings — e.g., `allow_session → kind="allow_always"` (Art4/L47), LLM as sole gate with no deterministic backstop (Art5/L46), `fadingThreshold >= freshThreshold` accepted without guard (Art3/L103)
- **D2 (Silent failure)**: 14 findings — e.g., bare `except Exception` swallowing all failures (Art4/L154), `detect_read_only_tools` returning `Vec<String>` with no `Result` (Art5/L122), NaN `updatedAt` propagating to `recall=1` (Art3/L140)
- **D3 (Missing contract guard)**: 12 findings — e.g., `relations` null crash on `for...of` (Art3/L97), `timeout <= 0` accepted without validation (Art4/L107), tool names used without sanitisation (Art5/L84)
- **D4 (Dead correctness code)**: 4 findings — e.g., `recall < 0` clamp unreachable (Art3/L72), `tokens.length === 0` guard after loop already exits (Art1/L7)

**Finding (taxonomy validity)**: All 49 structured-review findings classify into D1–D4 without residual. The taxonomy is exhaustive for this artifact domain. *Limitation*: D1–D4 was inductively derived from the same five artifacts used to demonstrate coverage; exhaustiveness holds within this set, not across held-out data. Future work should validate the taxonomy on an independent artifact corpus.

### 3.3 What CI Misses

These files exist in CI-integrated repositories; no defect was flagged by the project's automated checks (type-check, lint, test suite) at the time of review. We did not independently run CI pipelines — the claim is that the files were in committed, CI-integrated codebases, not that we executed those pipelines ourselves. The defects found in §6 were all present in files whose projects report passing CI. This is consistent with the structural argument: CI tests the syntactic envelope; the taxonomy captures what lies outside it.

---

## 4. Runtime-Decoupled Governance Gate (C2)

### 4.1 Architecture

The governance gate operates on the *files the generator writes*, not on the running program. This de-coupling is the key architectural property:

```
Generator output (files) → Governance gate → PASS / PENDING / BLOCKED verdict
                                                         ↓
                                                    CI pipeline
```

The gate reads files, applies the review protocol, and emits a structured verdict (PASS / PENDING / BLOCKED) with severity-graded findings (S/A/B). Because it operates on static artifacts, it requires no:

- Runtime environment
- API coupling to the generator

- Language-specific toolchain
- Access to the executing system

This makes it portable: the same gate that reviewed an OpenCode TypeScript permission module (Art1) also reviewed a Rust permission judge (Art5) and a Python callback (Art4) in the same session, with no tool-chain adaptation.

#### 4.2 Integration Contract

The gate is invocable as:

```
fh-gate review <artifact> [--verdict-only]
```

Output contract:

```
VERDICT: PASS | PENDING | BLOCKED
FINDINGS:
  [S|A|B] L{line} — {description}
```

The `PENDING` verdict on a CI-passing file is the primary signal: code that passed all automated checks has unresolved semantic defects. In our controlled trial (v2 companion paper, Experiment 2), `arity.ts` from OpenCode received a `PENDING` verdict with two A-grade findings despite passing the project's complete CI suite and developer self-review.

#### 4.3 Portability Evidence

Across the E-ablation artifact set (§6), the gate was applied to:

- TypeScript (Art1, Art3) — no TypeScript toolchain required
- Markdown skill specification (Art2) — no parser required
- Python (Art4) — no Python runtime required
- Rust (Art5) — no Rust compiler required

Five languages/formats, one protocol, one session. This is the operational definition of runtime-decoupled.

---

### 5. Methodology-as-Code (C3)

#### 5.1 The SKILL.md Pattern

The governance protocol is expressed as a `SKILL.md` file: a version-controlled, human-readable specification that an AI assistant can execute. The format is:

```
## Protocol
[Step-by-step review procedure]
```

```
## Axes
[Review dimensions with definitions]

## Output format
[Structured verdict template]

## Done when
[Explicit completion condition]
```

This is methodology-as-code in the SE sense: the review protocol is a first-class artifact, subject to version control, diff, review, and testing.

### 5.2 The Two-Layer Architecture

The FH governance layer has two tiers:

| Layer | Contents | Portability |
|---|---|---|
| **Methodology** | `SKILL.md` protocol, axes, output format, completion criteria | Model-agnostic, CLI-agnostic, version-controlled |
| **Automation** | Claude Code hooks, sub-agent dispatch, CI integration | Host-specific, CLI-dependent |

The methodology layer is portable: it can be injected into any LLM CLI (Claude Code, Codex, Gemini CLI, GitHub Copilot) as a plain-text prompt. The automation layer is host-specific but optional — the methodology layer works without it.

This separation has a practical consequence for SE teams: adopting the governance protocol requires no toolchain change, only a protocol document and any LLM access. The automation layer accelerates execution but is not a prerequisite.

### 5.3 Versioning and Evolution

The governance protocol is tracked in git. Changes to the protocol are subject to the same review process that reviews code. This is the self-referential property: the protocol that governs code changes is itself governed by the same process. Our companion paper (v2) reports that a steel-quench pass on the protocol itself found phantom references and structural gaps that the authors missed — an instance of the protocol finding defects in its own specification.

---

## 6. E-Ablation: Process vs. Capability (C4)

### 6.1 Experiment Design

**Motivation.** The central rival explanation for governance protocol effectiveness is: *any sufficiently detailed review prompt improves output, independent of harness structure.* If this is true, the governance protocol provides no unique value — a longer or more specific unstructured prompt would achieve the same coverage. E-ablation is designed to test this directly.

**Conditions:**

- **Condition A (Structured)**: FH governance protocol — 4 axes (design correctness / security / API contract / maintainability), 2-wave review (find → challenge/retract). Applied systematically to each axis before producing findings.
- **Condition B (Unstructured)**: "Find all design and security defects in this code. Grade S/A/B with line number." No axis specification, no multi-wave discipline.

Both conditions used the same model (Claude Sonnet 4.6, model ID `claude-sonnet-4-6`, 2026-06-02), same artifacts, and comparable prompt length. Independent agents were used per artifact to prevent cross-contamination within conditions.

**Ground truth.** A QA engineer with no prior exposure to the artifacts read all five independently and recorded observed issues before either AI condition ran. This establishes a human-perspective independent ground truth (GT) of 8 findings. The QA reviewer's background (process/quality focus, non-expert coder) represents a practitioner profile distinct from a security researcher, providing a complementarity-revealing GT.

**Artifacts.** Five AI agent permission and governance modules (100–219 lines each):

| # | File | Repo | Domain |
|---|---|---|---|
| 1 | `permission/arity.ts` | sst/opencode | Permission allowlist, AI-generated |
| 2 | `skills/opencode/SKILL.md` | NousResearch/hermes-agent | Agent orchestration spec |
| 3 | `memoryFreshness.ts` | tinyhumansai/openhuman | AI memory decay scoring |
| 4 | `acp_adapter/permissions.py` | NousResearch/hermes-agent | Dangerous-command approval bridge |
| 5 | `permission_judge.rs` | block/goose | LLM-based read-only classifier |

**6.2 Results**

**Finding counts:**

| | S | A | B | Total |
|---|---|---|---|---|
| Condition A | 3 | 19 | 27 | **49** |
| Condition B | 6 | 13 | 28 | **47** |

Total findings are nearly identical (49 vs 47), which indicates that the unstructured condition is not producing fewer findings overall. The distributions differ: Condition B assigns more S-grades (6S vs 3S) while reporting fewer A-grade issues (13A vs 19A). We previously read this differential as severity inflation by the unstructured condition. Blind and cross-family re-grading (§6.6) does not support that reading — both conditions over-graded against a blind rubric — and we therefore do not report the S-count differential as a result.

**GT coverage:**

| GT ID | Description | Condition A | Condition B |
|---|---|---|---|
| GT-1-1 | 138-entry hard-coded block, no structure or tests | PARTIAL | PARTIAL |
| GT-4-1 | Broad exception handling masks failures | ✅ HIT | PARTIAL |
| GT-5-1 | Deep if-nesting in permission extraction function | ⚡ PARTIAL | ⚡ PARTIAL |
| GT-2-1 | Document appears truncated mid-section | ❌ MISS | ❌ MISS |
| GT-2-2 | Code block formatting inconsistent | ❌ MISS | ❌ MISS |
| GT-2-3 | . . . placeholders make code non-executable | ❌ MISS | ❌ MISS |
| GT-3-1 | Dead code guards in recall formula | ✅ HIT | ❌ MISS |
| GT-3-2 | Three independent guards for same boundary | ⚡ PARTIAL | ⚡ PARTIAL |

**GT recall:**

| | Strict (full hits) | Lenient (full + partial) |
|---|---|---|
| Condition A | 2/8 = **25%** | 5/8 = **62%** |
| Condition B | 0/8 = **0%** | 4/8 = **50%** |

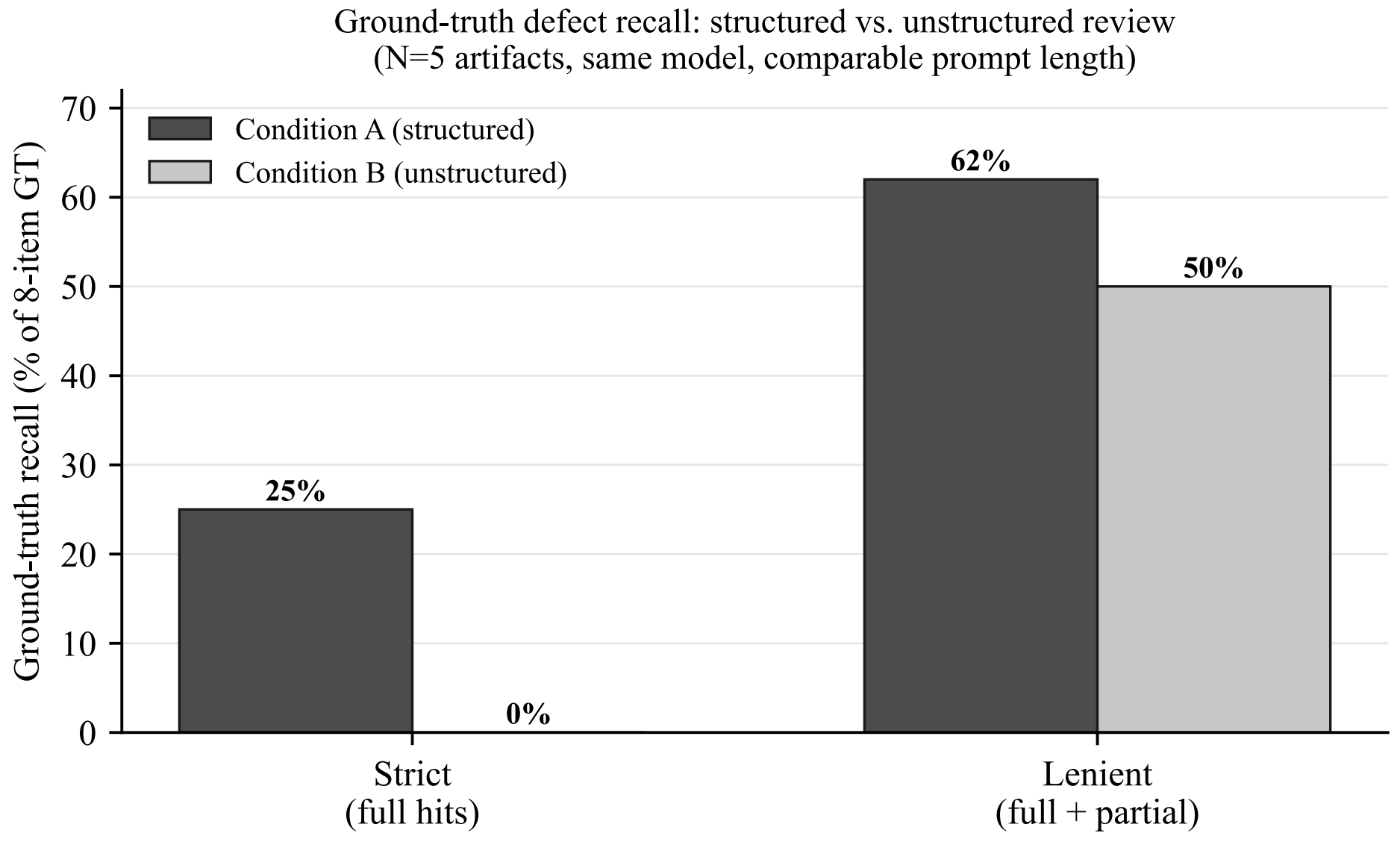

*Figure 2. Ground-truth recall: structured vs unstructured review.*

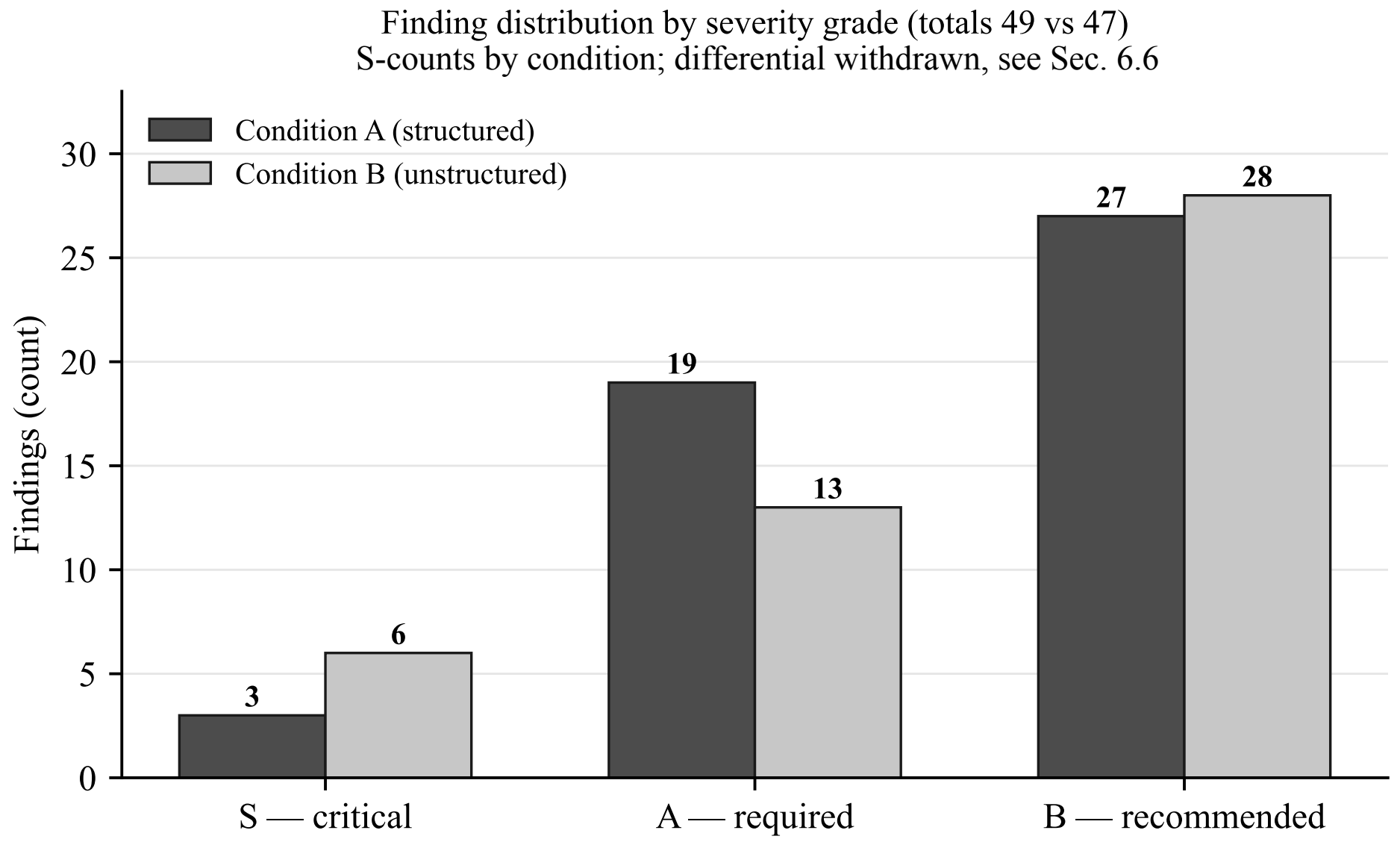

*Figure 3. Severity-grade distribution by condition.*

**6.3 Analysis**

> ***Session contamination caveat (L1).*** *Both conditions ran in the same AI session. Condition A (structured) ran first; Condition B may have been influenced by proximity to Condition A findings. All directional claims below should be interpreted with this limitation in view. Independent-session replication has since been performed and is reported in §6.6; it changes the finding-count and severity results; the recall result is re-tested directly in §6.7.*

**Strict recall in this run is concentrated in two findings.** Condition A's advantage on strict GT recall (25% vs 0%) in this run comes from two findings — and §6.7 reports that the unstructured condition finds both of them under independent sessions, so this concentration is a property of this run, not an exclusive capability of the structured protocol: GT-3-1 (dead code, caught by Axis 4 — maintainability) and GT-4-1 (exception handling, caught by Axis 2 — security). The structured Axis 4 explicitly instructs the reviewer to look for dead code and silent failure modes, which is a plausible mechanism for the gap in this run; it is not an exclusive capability, since §6.7 finds the unstructured condition covering both classes under independent sessions.

**Severity grades differed between conditions in this run.** Condition B escalates the same finding to S where Condition A grades it A or B: `required: []` in the Goose permission schema is [B] in Condition A, [S] in Condition B. `hardcoded model string` is [B] in Condition A, [S] in Condition B. We first read this as the unstructured condition applying idiosyncratic severity without an explicit grade calibration. That reading does not hold: under condition-blind re-grading the S-count differential does not survive and is withdrawn (§6.6).

**Both conditions miss document-quality defects.** GT-2-1~3 (document truncation, formatting inconsistency, non-executable placeholders in Art2/SKILL.md) were found exclusively by the human QA reviewer. Both AI conditions focused on runtime correctness and security, systematically bypassing spec-completeness and document-quality issues. This is complementary coverage in this sample rather than a failure: on these eight items the human reviewer's hits and the AI conditions' hits do not overlap, so the union exceeds either alone here. Whether the defect classes are non-overlapping in general is not established by five artifacts.

**Non-overlapping coverage within conditions.** Condition A found issues Condition B missed (Art3 `Infinity` halfLife disabling stale detection L102; Art2 session_id field unspecified L80). A third item previously listed here — a claim about the mutability of the Art1 permission table at L163 — is **retracted**. Under the re-grading reported in §6.6, a claim concerning L163 was overturned by direct source inspection, which found the table declared `const` without `export` and therefore module-private. The record does not resolve unambiguously which condition's phrasing of the L163 claim was the one overturned, so we withdraw the item from this non-overlap comparison rather than reassign it to the other condition. Condition B found issues Condition A missed (Art1 $O(n^2)$ complexity L2-6; Art3 ms/seconds unit confusion L140; Art5 idempotent annotation contradiction L64). This demonstrates that neither condition strictly dominates the other on total coverage, but their overlap profile differs: Condition A's unique finds are predominantly D1–D2 (semantic boundary, silent failure); Condition B's unique finds are predominantly performance and annotation issues.

### 6.4 Addressing the Rival Explanation

The rival explanation — "any detailed prompt finds the same issues" — predicts that Conditions A and B, given similar token counts, should find similar findings with similar GT recall. The results do not support this:

- GT strict recall in this run: 25% vs 0% — but see §6.7, where an independent-session re-test puts the two conditions one strict hit apart and this bullet does not survive

- Unique findings: 15–20 non-overlapping per condition (not a strict superset relationship)
- Grading consistency: in this run A assigned fewer S-grades than B, but this S-count differential does not survive blind re-grading and is withdrawn (§6.6)

**Caveat.** Both conditions ran in the same AI session, creating potential confirmation bias. The structured condition ran first; Condition B may have been influenced by proximity. Fully independent sessions (separate API invocations, no shared context) were subsequently run; their results, and what they do and do not settle, are reported in §6.6 and recorded in Limitation L1 (§8).

---

### 6.5 E-baseline: SAST Comparison

To test whether the taxonomy classes overlap with what static analysis already catches, we ran SAST tools against the same five artifacts and compared their output to the ground-truth findings. Semgrep 1.164.0 (rulesets `p/default`, `p/python`, `p/typescript`, `p/javascript`, `p/rust`, `p/security-audit`) scanned the four code artifacts with no parse errors and produced **zero findings**. Ruff with every rule enabled (`--select ALL`) flagged the Python artifact's broad `except Exception` (BLE001) and a set of import/docstring/annotation style issues; no other tool flagged any ground-truth finding.

Finding by finding: the dead/unreachable guards (D4) were missed — they are unreachable only given a bounded computation (`2^(-age/halfLife) ∈ (0,1]`), which no SAST rule derives; the inconsistent three-site defensive layering (D3) and the hard-coded-table maintainability issue were missed as cross-site design observations; the deep-nesting complexity finding was missed (the relevant complexity lint is opt-in). The single point of contact was the broad-`except` pattern (D2): Ruff's BLE001 flags the *syntactic pattern* but not the *semantic* finding — that the swallowed failure modes are indistinguishable to the caller. Ruff's remaining output was style noise absent from the governance taxonomy.

On this set, the design- and semantic-defect classes (D1–D4) are therefore **empirically non-overlapping** with default SAST coverage: SAST and governance review target different defect spaces. This is the measured form of the structural argument (§3) — and, if anything, stronger than we predicted: SAST missed even the D4 dead-code finding, because *statically-unreachable* and *mathematically-unreachable* are not the same thing. *Caveat:* this is one curated SAST configuration; a bespoke custom-rule effort could encode some of these patterns by hand — which is precisely the manual design-encoding the governance methodology automates as review.

---

### 6.6 Replication

Two replications were run after the primary experiment. The first addresses the session-context confound recorded as L1; the second addresses a confound the first leaves untouched, namely that the grading of all findings was performed by a single party who knew each finding's condition.

**Cross-session replication.** Conditions A and B were re-run over the same five artifacts in ten fully independent agent sessions, five per condition, with no shared context between any two sessions and no access to the results of the original run. Finding counts by artifact:

| Artifact | Condition A (S/A/B) | Condition B (S/A/B) |
|---|---|---|
| `arity.ts` | 0/3/5 = 8 | 4/7/3 = 14 |
| `SKILL.md` | 1/3/6 = 10 | 4/6/5 = 15 |
| `memoryFreshness.ts` | 0/3/5 = 8 | 2/6/4 = 12 |
| `permissions.py` | 0/4/5 = 9 | 3/5/4 = 12 |
| `permission_judge.rs` | 3/4/3 = 10 | 3/5/3 = 11 |
| **Total** | **4/17/24 = 45** | **16/29/19 = 64** |

The direction of the raw count reverses. In the same-session run the structured condition produced marginally more findings than the unstructured one (49 vs 47); under independent sessions it produced substantially fewer (45 vs 64), and the reversal holds at A-grade as well (17 vs 29). The S-count gap moves the other way and widens (4 vs 16).

Two consequences follow, and the second is the more serious. First, raw finding count is not stable across the two designs and cannot be read as a coverage measure in either direction. Second, and this is what the replication does not resolve: **ground-truth recall was not recomputed under the independent-session design.** The recall figures reported in §6.2 remain measurements of the original same-session run — the run L1 identifies as confounded. The replication removes the confound for the finding-count and severity-distribution results, and leaves the recall result, which carries this paper's central comparative claim, measured only under the confounded design. §6.7 re-measures it.

**Blind re-grading.** The findings of the original run were pooled per artifact, deduplicated, shuffled, and stripped of both their original grade and their condition label. Five isolated graders, one per artifact, regraded the pool working from the locked artifact source, the artifact's ground-truth items, and a fixed rubric: S for an exploitable or fail-open defect that blocks normal operation, A for a real but non-blocking defect, B for a minor issue, and FP for a claim that is factually wrong about the code. No grader saw the experiment design, the conditions, or any prior grade.

The pooled re-grade assigned **zero S on all five artifacts**: 0S, 21A, 46B, and 8 findings judged false positives, across approximately 75 deduplicated items. The seven distinct S-grades of the original run — three from Condition A, six from Condition B, two of which both conditions assigned independently — were reassigned without exception:

| Original S-grade | S-origin | Blind verdict |
|---|---|---|
| Art4 L130 `allow_permanent` default `True` | both conditions | A |
| Art5 L46 LLM as sole permission gate | both conditions | A |
| Art4 L47 `allow_session` to `allow_always` | Condition A | B |
| Art1 L163 re-export | Condition B | A |

| Original S-grade | S-origin | Blind verdict |
|---|---|---|
| Art2 L186 hard-coded model string | Condition B | B |
| Art5 L97 empty read-only allowlist | Condition B | **FP** |
| Art5 L63 `required: []` | Condition B | B |

**Cross-family re-grading.** Because all five blind graders belonged to the same model family as the reviewing model, the blind re-grade was repeated on the same condition-stripped pool, deduplicated to 69 items, using two different model families and one human rater:

| Grader | S | A | B | FP | Items |
|---|---|---|---|---|---|
| Claude (blind) | 0 | 21 | 46 | 8 | ~75 |
| Gemini 0.41 | 0 | 30 | 32 | 7 | 69 |
| GPT-5.5 | 0 | 23 | 32 | 14 | 69 |
| Human rater (7 S-items only) | 0 | — | — | 0 | 7 |

All three model families reach zero S independently. The human rater, given the seven disputed S-items with source snippets, initially retained one — the claim that the Art1 permission table is externally mutable — and withdrew it after checking the source, where the table is declared `const` without `export` and is therefore module-private. The withdrawal falsifies a claim that appears in the original per-condition findings and was cited in §6.3 as a non-overlapping find; we have retracted it there. Two properties of this arm must be stated plainly. The human rater is the same person who produced the researcher ground truth of §6.1 and is not independent of the author team, so this arm is not external validation; it is a self-refutation reached by source inspection, and should be read only as that. And the human graded from a summary table with source snippets rather than from full source, so fail-direction tracing was out of reach by construction; the arm bears on the S-retention threshold and not on false-positive detection.

**What this withdraws.** The severity reading stated in §6.2 — that Condition B's higher S count indicates unstructured over-grading relative to Condition A — does not survive. Under a blind rubric applied by graders of three families, both conditions over-graded: Condition A's own three S-grades became two A and one B, and the two S-grades the conditions agreed on became A. The defensible statement is not that the structured condition graded more accurately, but that **absolute S-counts are not a robust measurement**: they survive neither a change of grader nor a change of grader family. We withdraw the S-count differential as a reported result.

A second claim is withdrawn on the same ground. The same-family blind re-grade found its eight false positives skewing toward Condition B, which would support a precision reading in relative form. That skew does not replicate: one of the other families produced false positives leaning slightly toward

Condition A, and the other split them evenly. The false-positive-rate comparison is a property of one grader family, not of the conditions, and we do not report it.

**What remains, and how weak it is.** One finding the unstructured condition graded S is a false positive on the code: a parse failure in Art5 yields an empty read-only allowlist, which auto-approves nothing and therefore fails closed rather than open. This was reached independently by an Opus second-wave challenge, by the same-family blind grader, and by Gemini, with GPT-5.5 concurring at B — four determinations across three families.

Beyond that single item, one directional regularity holds across every grader: no Condition-A S-grade was ever judged a false positive, whereas every S-to-FP reassignment, in every grader, originated in a Condition-B S-grade. We report this as directional only, and attach its base explicitly. Of the seven distinct S-grades, exactly one was unique to Condition A and four were unique to Condition B; the asymmetry is therefore computed over one structured-only item against four unstructured-only ones. That is not a base from which a magnitude can be stated, and we state none.

Ground-truth recall is unaffected by re-grading in principle, since re-grading reassigns severity rather than changing which ground-truth items were matched. We note two reasons not to lean on this. The blind re-grade marked eight pooled findings as factually wrong about the code, including one in Art3 and one in Art4 — the two artifacts that supply Condition A's only two strict ground-truth hits — and the record does not identify which items those were, so the possibility that a credited hit was among them is not excluded. Recall was not recomputed in the replications reported in this section. It was re-measured afterwards in an independent session under condition-blind scoring, and that re-test does not reproduce the contrast (§6.7). The recall result therefore remains the claim on which this paper's conclusion rests, and the one replication that has since tested it did not replicate it.

---

## 6.7 Recall Re-Test Under Independent Sessions

The replication of §6.6 removed the session-context confound for the finding-count and severity results but left the recall result — this paper's central comparative claim — measured only under the confounded design. We re-measured it. The result is reported here descriptively; it is small, and it does not support a superiority claim in either direction.

**Design.** The five artifacts were re-fetched at the locked commits of §6.1 (line counts 163 / 219 / 174 / 168 / 162, matching the original; file hashes recorded with the run). Both conditions were re-run on the same model identifier as the original experiment, `claude-sonnet-4-6`, every run in a fully independent session with no shared context. Each of the eight ground-truth items belongs to one artifact, so a single pass over the five artifacts scores all eight; we ran three such passes per condition, giving 24 item-opportunities per condition and 15 reviews per condition in total. The structured condition was supplied the protocol document itself. Scoring was performed by a **single blinded grader** of a different model family, which received the ground-truth items and one review at a time with no knowledge of the producing condition, and returned HIT / PARTIAL / MISS under the §6.2 definitions. Two fixtures were passed through it first — a verbatim restatement of two ground-truth items, and two unrelated observations — which it scored HIT/HIT and MISS/MISS. That is a discrimination check, not a validation of the HIT/PARTIAL boundary, and there is no second grader and therefore no inter-rater figure.

| Condition | Strict hits (3 replicates × 8 items) | Lenient | §6.2 single-session |
|---|---|---|---|
| A — structured | 6 / 24 | 11 / 24 | 2/8 strict, 5/8 lenient |
| B — unstructured | 5 / 24 | 8 / 24 | 0/8 strict, 4/8 lenient |

**What this shows.** The contrast reported in §6.2 — 25% strict against 0% — does not replicate. Under independent sessions and blind scoring the two conditions differ by **one strict hit across 24 opportunities**. The two findings §6.3 attributes the structured advantage to, the dead-code guard and the broad exception handler, are both found by the unstructured condition here. The structured condition's own strict rate, 6/24, is close to the 2/8 of §6.2; what moved is the unstructured condition, from 0/8 to 5/24. A single-session, single-replicate 0% was therefore within the range that repeated independent draws produce, and should not have been read as a floor. The three document-quality items (GT-2-1 to GT-2-3) remained missed by both conditions in every replicate on this artifact set.

**Where the two-item contrast came from.** The zero is a property of all eight items — none was graded a full hit — so no two items account for it. What two items do account for is the entire strict gap between the conditions, since those are the only items on which the structured condition scored full hits. On GT-3-1 the §6.2 record reports nothing from the unstructured condition at the dead-guard site. On GT-4-1 it reports the correct line described as a redundant exception clause rather than as broad handling that masks failures, and records PARTIAL on that basis, against HIT for the structured condition, which named the masking mechanism at the same line. We state these as what the record reports rather than as adjudicated facts, for the reason given below. In the re-test the unstructured condition reports the masking mechanism and finds the dead guard in every replicate. What changed is therefore the review output, not only the grading — which is consistent with, though it does not establish, a change in what the model identifier serves. The §6.2 grading itself was performed by one party who knew each finding's condition, and no adjudication record exists for it; the re-test's grading was blind for that reason.

**A prompt-length control.** The unstructured condition was run twice, once with the single-sentence prompt of §6.1 and once padded to comparable length as the original design specifies. Both give 5/24 strict, so prompt length does not account for the difference from §6.2.

**An unexplained level shift.** Yield per artifact-review is roughly one third of the original replication's on the same artifacts, same model identifier and same independent-session design: structured 0.8 against 1.8 findings per artifact-review, unstructured 1.6 against 2.6 with the length-matched prompt. Prompt length accounts for part of the gap and the remainder is unexplained. We did not verify that the model served under a given identifier is unchanged across three months, and we have no means to; a stable model identifier is not a stable measurement instrument. This is why the re-test is reported as a same-day comparison between conditions rather than as a correction of the §6.2 figures.

**A packaging observation, reported as exploratory.** The re-test was first run with the structured condition expressed as a two-sentence paraphrase of §6.1's description rather than as the protocol file. That arm scored 3/24 strict — below the unstructured condition. The paraphrase was neither semantically nor token matched to the document, so this is not a controlled comparison and it does not establish anything about methodology-as-code; it motivates a factorial test separating protocol content, packaging and prompt length, which we have not run.

**Scope.** Three replicates per condition, eight ground-truth items, one grading family, one model identifier, no randomisation of run order and no reported control of runtime settings beyond the model identifier. The re-test is here because it closes the gap §6.6 left open, not because it settles the question. Per-item and per-replicate scores are available from the author on request.

### 6.8 Pre-Registered Test: Model Tier × Review Method

**Design.** A 2×2 test, pre-registered and sealed before any run: two model tiers of one family (Claude Haiku 5.5 and Claude Opus 5.5) × two review methods (the structured protocol of §6.1, about 54 KB, against the one-sentence unstructured prompt of §6.1). Targets were the five E-ablation artifacts plus eight files with publicly disclosed security defects (16 ground-truth items in all) and seven post-fix control files; case files ran three times per cell, controls once — 184 runs. Grading was done by a model from a different family, blind to condition; lenient = HIT or PARTIAL, strict = HIT; intervals are 95% bootstrap over ground-truth items.

| Scoring | Method effect (structured − one sentence) | Tier effect (Opus − Haiku) | Pre-registered verdict |
|---|---|---|---|
| Lenient (primary) | −7.3 pp [−14.6, −1.0] | +5.2 pp [−2.1, +13.5] | cannot decide |
| Strict (secondary) | −6.3 pp [−15.6, +2.1] | +16.7 pp [+7.3, +26.0] | tier larger in this sample |

The primary verdict is "cannot decide" not because both effects sit near zero but because the method effect left zero in the negative direction, for which the pre-registered rule had no verdict. Under strict scoring only the tier effect, of the two main effects, excludes zero. A secondary comparison, not used for the verdict: under strict scoring the small model with the structured prompt scored below the large model with the one-sentence prompt (−22.9 pp [−39.6, −8.3]; lenient −12.5 pp [−25.0, 0.0]).

**Follow-ups.** Repeating the design with the previous model generation (Claude Haiku 4.5, Claude Opus 4.6; 184 further runs) again found no lead for the structured prompt (lenient −3.1 pp [−15.6, +10.4]) and a tier effect that excluded zero (lenient +13.5 pp [+3.1, +25.0]); the pre-registered comparison, the difference in method effect between generations, was +4.2 pp [−10.4, +20.8], cannot decide. Running both methods on an older and a current release of the agent CLI with two Haiku generations (366 of 368 runs completed; a context-contamination check in that environment was later found unreliable) did not show a larger structured share on the older release (lenient −4.4 pp [−15.1, +6.2]; cannot decide).

**Limits.** The method contrast confounds structure with length (about 54 KB against one sentence), so it does not separate the two. Three of the 16 ground-truth items scored zero in every cell and a fourth in all but one, leaving about 12 informative items. One model family, two tiers.

## 7. Supporting Evidence (Shared with Companion Paper)

Three experiments from our companion empirical paper provide supporting evidence for the cs.SE thesis. We summarize them without reproducing the full analysis, which belongs to the companion paper.

**E-trial (Controlled governance verdict flip)**: `packages/opencode/src/permission/arity.ts` — 163 lines of AI-generated TypeScript, CI-passing, developer-approved. Under FH governance review: verdict PENDING, two A-grade findings (short-token overflow pattern; five common AI-CLI tools absent from the allowlist). Neither finding was surfaced by CI, linter, or developer self-review. This is an N=1 instance of the semantic gap between syntactic and design correctness.

**E-tier (Model tier comparison)**: On one bounded artifact (`pipeline-conductor/SKILL.md`, 417 lines), with one run per tier and severity grades assigned by the reviewing model itself (no independent answer key), Haiku-4.5, Sonnet-4.6, and Opus-4.8 each reported the same three S-grade findings under the same governance protocol. Opus additionally raised an architectural meta-critique that the other tiers did not. This single comparison does not test whether process outweighs capability; §6.8 does.

**E-panel (Multi-session coverage)**: Across N=5 FH-internal artifacts, a single-session review (C1) achieved 57% defect coverage. A three-persona cross-session panel (C2) raised coverage to 84%. The delta (27 percentage points) at negligible marginal cost supports the value of structured independent review passes.

---

## 8. Threats to Validity

**L1 — Same AI session for both E-ablation conditions (replicated; partially resolved).** Condition A ran before Condition B in the same session, so the reviewing model may have retained Condition A findings when generating Condition B findings. This was addressed by re-running both conditions over the same five artifacts in ten fully independent sessions, five per condition, with no shared context (§6.6). For the finding-count and severity-distribution results the confound is removed, and those results change: the structured condition produced fewer total findings than the unstructured one (45 vs 64) rather than marginally more (49 vs 47), and the reversal holds at A-grade (17 vs 29), while the S-count gap widened in the other direction (4 vs 16). The residual is specific and load-bearing. **Ground-truth recall was not recomputed under the independent-session design of §6.6. It has since been re-measured directly, under independent sessions with blind scoring, and is reported in §6.7: the 25%-against-0% contrast of §6.2 does not replicate — the two conditions come out one strict hit apart in 24. The §6.2 figures therefore stand as measurements of the confounded run and should** **not be read as a comparative result. The residual that remains is narrower and is stated in §6.7: three replicates, eight items, one grading family, one model identifier, and an unexplained three-fold drop in yield against the original replication under that same identifier.**

**L2 — Single human rater for ground truth. The 8-item GT was established by one QA engineer. Single-rater GT is subject to perspective bias (the rater's QA background shaped which defect classes were visible). Mitigation: inter-rater GT with a second independent reviewer; Cohen's kappa for agreement.**

**L3 — Small N (5 artifacts). E-ablation findings are based on N=5. Statistical power is insufficient for significance tests. All claims are directional; replication with N=20+ is needed for venue submission.**

**L4 — Artifact selection bias. Five artifacts were selected from AI agent governance code by the researchers. They may be atypical — permission and security code may have higher semantic defect density than other AI agent code. A random sample from a larger corpus (e.g., a random selection from GitHub AI agent repositories) would reduce selection bias.**

**L5 — Self-referential evidence. Art2 (Hermes SKILL.md) is itself a governance artifact of a kind related to FH. The governance methodology reviewing another governance methodology may have domain-specific advantages not present when reviewing unrelated code.**

**L6 — E-baseline configuration breadth. §6.5 reports an E-baseline against Semgrep and Ruff, finding the D1–D4 classes empirically non-overlapping with default SAST coverage. The residual limitation is breadth: a single curated SAST ruleset set was used, and standalone clippy (Rust) / eslint (TS) were not exercised. A maximally-tuned, multi-tool SAST sweep over a larger corpus would harden the non-overlap claim.**

**L7 — Model-based verdict provenance. The structured-review findings are themselves outputs of the reviewing model; the gate's accept/flag signal is a model judgment, which an adversarial input or the model's own optimism can corrupt. The current design mitigates this only by structure (explicit axes) and human-in-the-loop adjudication, not by binding the terminal verdict to non-model evidence. A stronger gate would anchor each finding to a literal span in a declared source (static side) or an observed execution result (dynamic side); we treat non-model verdict provenance as a robustness direction (§10).**

**L8 — Grade-boundary subjectivity. The S/A/B scale used throughout §6 is applied by the reviewing model and has no external calibration. Blind re-grading of the pooled findings, with condition and grade labels removed and graders working from the locked source and a fixed rubric, assigned zero S across all five artifacts; the result held for three model families and, after source verification, for a human rater (§6.6). Every one of the seven distinct S-grades in the original run was reassigned to A, B, or false positive. Part of the collapse is rubric calibration rather than error: the blind rubric reserves S for defects that block normal operation or are directly exploitable, whereas the original review admitted defects that require adversarial input. The conse-**

quence is therefore not that the original S-grades were wrong in fact, but that the S boundary is grader-dependent and absolute S-counts are not a stable quantity. Every comparison in this paper that rests on severity-grade counts is weakened accordingly, and we have withdrawn the two such comparisons that were previously reported (§6.6). Comparisons resting on ground-truth recall are not affected by grader calibration, though they carry the separate limitation recorded in L1. Making severity a reportable quantity would require a grade scale anchored to an external, source-checkable criterion rather than to reviewer judgment; we treat that as future work (§9).

---

## 9. Future Work

E-baseline at scale. §6.5 reports a first E-baseline (Semgrep + Ruff) on the five-artifact set, with the D1–D4 classes empirically non-overlapping with default SAST. Extending it — additional tools (clippy, eslint, CodeQL), a maximally-tuned ruleset, and a larger artifact corpus — would carry the measured non-overlap toward a venue-grade result.

Ground-truth recall under independent sessions. Run and reported in §6.7. The 25%-against-0% contrast of §6.2 does not replicate under independent sessions with blind scoring; the two conditions come out one strict hit apart in 24. What remains open is replication at a size that can separate a one-hit difference from noise, a second grading family for an inter-rater figure, and an account of the threefold yield drop observed under an unchanged model identifier.

Larger N. Expand to N=20 artifacts across diverse domains (API specs, CI configs, auth middleware, database schemas). Stratify by artifact type and measure within-stratum GT recall.

Developer study. Present the same five artifacts to a sample of developers (N=10–20) without AI assistance. Compare their defect recall against Conditions A and B. This would establish whether the governance protocol outperforms human expert review, not just unstructured AI review.

Inter-rater GT. Replicate the pre-read GT with two independent raters from different backgrounds (security engineer + QA engineer). Measure inter-rater agreement and use the intersection as a conservative GT.

Dynamic complementary layer. The governance gate evaluated here is static: it reads the artifacts the generator writes and reasons over their text. A structural bound limits this reach. Non-trivial semantic properties of programs are undecidable in the general case (Rice's theorem), and a further class of defects depends on *realized execution state* — values, timing, environment, and interaction order that are absent from the source artifact. Dynamic verification therefore adds coverage for defects whose manifestation depends on realized runtime state. A complementary observational layer that executes generated artifacts and checks realized behaviour is what would extend coverage to the realized-state class; it is a distinct line of work from the semantic-static gap (D1–D4) addressed here, and we are careful not to claim coverage of the realized-state class.

Multi-modal review. The taxonomy and gate evaluated here operate on a single textual artifact. Requirements, however, span modalities — natural-language specifications, visual mockups, API contracts — and a class of defects lives in *cross-modal inconsistency*: a requirement stated in text that contradicts its visual realization, or an API contract the implementation does not honour. A concrete realization pairs a text reviewer (extracting requirements from a specification) with a visual reviewer (inspecting the corresponding design artifact) and cross-validates the two, surfacing mismatches that neither modality flags alone. Extending the governance methodology to cross-modal review is a distinct direction beyond the single-modality D1–D4 set evaluated here (cf. multi-modal fact verification for LLMs, Patel 2025); we flag it as future work and do not claim a cross-modal contribution in this paper.

---

## 10. Related Work

Code review effectiveness. Fagan (1976) established that structured inspection yields higher defect detection rates than unstructured walkthrough. Our E-ablation asks whether that result extends to AI-conducted review. The single-session run of §6.2 is consistent with it; the independent-session re-test of §6.7 is not, resolving no difference between the two conditions. We therefore report the question as open for the LLM context rather than as an extension of the human inspection literature.

LLM-for-SE. Recent work applies LLMs to code review, bug detection, and vulnerability scanning (Pearce et al. 2022). These works treat the LLM as the primary agent. Our contribution is orthogonal: we treat the LLM as a *substrate* for executing a methodology, and attempt to separate the methodology's effect from the substrate — an attempt this paper does not complete, since all conditions ran on a single model identifier.

LLM-based test generation and oracles. A growing line applies LLMs to test generation and the oracle problem. APITestGenie (Pereira et al. 2024) generates web-API tests from requirements and OpenAPI specifications using retrieval-augmented prompting; LLM-Explorer (Zhao et al. 2025) drives cost-efficient mobile-app exploration by maintaining LLM-held knowledge rather than invoking the model per action. For the oracle problem specifically, Xu et al. (2025) generate JUnit test oracles by multi-LLM panel consensus, while DLLens (2024) synthesises differential-testing counterparts for deep-learning libraries. A pattern across the consensus-based line is that the terminal acceptance signal is itself a model judgment — an oracle agreed on by reasoning models rather than anchored to a non-model observation (an executed result, or a literal span in a declared source); differential approaches are notable precisely for re-anchoring the oracle in a non-model comparison. Our methodology is orthogonal to these test-*generation* systems: it contributes a taxonomy of the design-level defects such pipelines do not target (§3) and, as a robustness direction (§8, L7), the discipline of binding a gate's verdict to non-model evidence rather than a model's self-assessment. Such a discipline can be realized concretely by layering a model-consensus oracle — for instance a rotating-adjudicator ensemble with majority, plurality, unani-mous, and weighted voting and an explicit consensus score — with a *non-model backstop* that rejects structurally invalid outputs before the consensus is trusted, blocking the failure class in which several models agree on the same plausible-but-wrong structure.

Natural-language harnesses. Pan et al. (2026) study the agent harness itself as an executable natural-language object and ablate its components empirically (§2.4). That work and the present paper independently arrive at the same artifact form — harness control as natural-language specification rather than code — from different communities and toward different ends: NLAH measures the form's effect on task resolution, while this paper treats the specification as a governed software artifact and measures defect coverage. The two are complementary rather than overlapping.

AI agent safety and permission systems. Work on tool-use safety and indirect prompt injection (Greshake et al. 2023) motivates the domain of our artifact set. Our taxonomy (D1–D4) provides a structured vocabulary for the defect classes these systems are designed to prevent.

Methodology-as-code. Infrastructure-as-code (Hashicorp Terraform, Pulumi) established that operational procedures can be version-controlled software artifacts. Our SKILL.md pattern applies the same principle to review protocols: the procedure is code, subject to the same development lifecycle as the software it reviews.

Static analysis baselines. Semgrep (semgrep.dev), CodeQL (Avgustinov et al. 2016), and Infer (Calcagno et al. 2015) represent the state of the art in automated semantic defect detection. Our E-baseline (§6.5) measures the non-overlap between these tools and governance review on our artifact set: on the five artifacts, default SAST coverage and the D1–D4 design/semantic classes did not overlap.

---

## 11. Conclusion

This paper argues that a structured governance methodology layer addresses a class of defects in AI-assisted software development that automated pipelines cannot reach: semantic boundary violations, silent failure modes, missing contract guards, and dead correctness code. We grounded this claim in a five-artifact external evaluation drawn from AI agent permission and governance code, establishing independently collected human labels as ground truth and running a controlled ablation comparing structured and unstructured AI review on the same artifacts.

The E-ablation run of §6.2 records higher ground truth recall for structured review (62% vs 50% lenient, 25% vs 0% strict) than for token-matched unstructured review. That contrast does not replicate: under independent sessions with blind scoring the two conditions are one strict hit apart in 24 (§6.7). What survives is that structured review's own rate is stable across designs while the unstructured 0% was a single draw. We previously also reported more consistent severity grading; that reading does not survive the blind and cross-family re-grading of §6.6 and is withdrawn. The recall figures themselves were measured under the single-session design of §6.1;

§6.7 re-measures recall under independent sessions with condition-blind scoring, where the structured condition's 25% strict recall reproduces in all three replicates and the unstructured condition's 0% does not (it scores 21%). The rival explanation — that any detailed prompt achieves equivalent coverage — is not excluded by the data; the re-test resolves no difference between the two conditions at this sample size.

Both conditions miss document-quality defects identified by a human QA reviewer, with the two sets of hits not overlapping on these eight items, so a team using both covers more of this sample than either alone.

Position. We take model capability to set the ceiling: in our own pre-registered test the structured prompt raised recall under neither scoring, while under strict scoring the model-tier effect excluded zero (§6.8). What a harness changes is how much of that ceiling is delivered reliably. With the model held fixed, a change to the harness alone has been reported to raise a coding agent's Terminal Bench 2.0 score by 13.7 points (Trivedy 2026). On one incident-derived task that we authored, the side obligation was missed in all six runs (three per tier, Haiku and Sonnet), and neither a self-verification prompt nor resident instruction text without hooks or gates changed this (0/6 each); the original incident was caught when the repository state was inspected, outside the model. We state as a position, not as a measured effect of this paper, that the structure around a model is a separate axis from its capability, and that this structure is the harness.

The methodology is expressed as version-controlled, executable, CLI-portable artifacts (SKILL.md), decoupled from the runtime environment of the code being reviewed. This portability makes it wrappable around any code-generation tool without API coupling.

We release the defect taxonomy, the gate protocol, and the ablation prompts as supplementary material; per-replicate scores and the appendix tables are available from the author on request (§6.4, App. A, App. B). A first E-baseline (§6.5) finds the design/semantic defect classes empirically non-overlapping with default SAST; a multi-tool sweep at larger N is the next empirical step before venue submission.

---

---

## Appendix A — E-ablation Raw Findings

Full per-artifact findings for Conditions A and B are available from the author on request.

## Appendix B — Ground Truth Record

The full ground-truth record with rater notes is available from the author on request.

## Appendix C — Artifact Version Hashes

| Artifact | Repository | Commit fetched |
| --- | --- | --- |
| `arity.ts` | anomalyco/opencode | dev branch, 2026-06-02 |
| `opencode/SKILL.md` | NousResearch/hermes-agent | main, 2026-06-02 |
| `memoryFreshness.ts` | tinyhumansai/openhuman | main, 2026-06-02 |
| `permissions.py` | NousResearch/hermes-agent | main, 2026-06-02 |
| `permission_judge.rs` | block/goose | main, 2026-06-02 |